\documentclass[twocolumn,preprintnumbers,amsmath,amssymb]{revtex4}

\usepackage{graphicx}
\usepackage{dcolumn}
\usepackage{bm}
\usepackage{multirow}
\usepackage{color}
\usepackage[normalem]{ulem}
\usepackage{amsmath}
\usepackage{gensymb}
\usepackage[colorlinks=true]{hyperref}

\begin{document}

\title{Unveiling the optical emission channels of monolayer semiconductors\\
coupled to silicon nanoantennas}
\author{Jean-Marie Poumirol$^1$}
\author{Ioannis Paradisanos$^2$}
\author{Shivangi Shree$^2$}
\author{Gonzague Agez$^1$}
\author{Xavier Marie$^2$}
\author{Cedric Robert$^2$}
\author{Nicolas Mallet$^3$}
\author{Peter R. Wiecha$^3$}
\author{Guilhem Larrieu$^3$}
\author{Vincent Larrey$^4$}
\author{Frank Fournel$^4$}
\author{Kenji Watanabe$^5$}
\author{Takashi Taniguchi$^6$}
\author{Aur\'elien Cuche$^1$}
\author{Vincent Paillard$^1$}
\email{vincent.paillard@cemes.fr}
\author{Bernhard Urbaszek$^2$}
\email{urbaszek@insa-toulouse.fr}

\affiliation{\small$^1$CEMES, Universit\'e de Toulouse, CNRS, Toulouse, France}
\affiliation{\small$^2$LPCNO, Universit\'e de Toulouse, INSA-CNRS-UPS, 135 Avenue Rangueil, 31077 Toulouse, France}
\affiliation{\small$^3$LAAS, Universit\'e de Toulouse, CNRS, Toulouse, France}
\affiliation{\small$^4$CEA-LETI, Universit\'e Grenoble-Alpes, Minatec Campus, Grenoble, France}
\affiliation{\small$^5$Research Center for Functional Materials, National Institute for Materials Science, Ibaraki, Japan}
\affiliation{\small$^6$International Center for Materials Anorthite, National Institute for Materials Science, Ibaraki, Japan.}

\begin{abstract}
Monolayers (MLs) of transition metal dichalcogenides (TMDs) such as WSe$_2$ and MoSe$_2$ can be placed by dry stamping directly on broadband dielectric resonators, which have the ability to enhance the spontaneous emission rate and brightness of solid-state emitters at room temperature. We show strongly enhanced emission and directivity modifications in room temperature photoluminescence mapping experiments. By varying TMD material (WSe$_2$ versus MoSe$_2$) transferred on silicon nanoresonators with various designs (planarized versus non-planarized), we experimentally separate the different physical mechanisms that govern the global light emission enhancement. For WSe$_2$ and MoSe$_2$ we address the effects of Mie Resonances and strain in the monolayer. For WSe$_2$ an important additional contribution comes from out-of-plane exciton dipoles. This paves the way for more targeted designs of TMD~-Si nanoresonator structures for room temperature applications. 

\end{abstract}

\maketitle

\textbf{Introduction.--- } Integrated efficient quantum light sources need to be developed for different applications including quantum computing and transmitted or stored information for cryptography. Key technologies are plasmonic or dielectric nanostructures (NS) coupled to various quantum emitters positioned in the NS near-field. Both the emitting dipole position and orientation with respect to the nanoantenna are important to optimize the emission rate and the emitted light directivity and waveguiding. Therefore, monolayers of semiconducting transition metal dichalcogenides (TMDs)  \cite{mak2016photonics,wang2018colloquium} such as WSe$_{2}$ and also heterobilayers \cite{nagler2019interlayer} are very promising as an active medium. Their atomic-monolayer thickness makes them uniformly sensitive to the near field around a nanoantenna, and they host several kinds of emitting dipoles at different wavelengths, such as the bright exciton (X$_{o}$, dipole in the monolayer plane), the so-called dark exciton (X$_{D}$, dipole out-of-plane) and other identified complexes such as charged excitons, localized excitons and biexcitons \cite{zhou2017probing,PhysRevLett.119.047401,barbone2018charge}. TMDs coupled to plasmonic NS have shown strongly enhanced emission and tailored directivity \cite{wang2016giant,kern2015nanoantenna,najmaei2014plasmonic}. However, plasmonics is limited by high losses in metals, heating and incompatibility with complementary-metal-oxide-semiconductor (CMOS) fabrication processes \cite{won2019into}. High refractive index dielectric NS such as  silicon-NS or GaP-NS \cite{wiecha2017evolutionary,cihan2018silicon,sortino2019enhanced}, can also provide reduced mode volumes, but in contrast to plasmonics they show low non-radiative losses and are compatible with large-scale semiconductor fabrication processes \cite{bidault2019dielectric,kuznetsov2016optically}. 
The emission properties of excitons in TMDs placed on NS have been investigated for several combinations of NS geometry and TMD materials \cite{cihan2018silicon,sortino2019enhanced,zhang2017unidirectional,chen2017enhanced,lee2017single,duong2018enhanced,bucher2019tailoring}. 
Despite these promising first results, the physical mechanisms that govern changes in the TMD emission as they are placed in proximity of the nanoresonators need to be understood. In addition to the targeted modification of the local electric field due to the resonator, several other mechanisms such as strong local strain induced by the resonator  \cite{sortino2020dielectric} and changing optical dipole orientation with respect to the detection contribute to brightening and change in directivity of the emission. \\
\begin{figure*}[t]
\includegraphics*[width=0.85\linewidth]{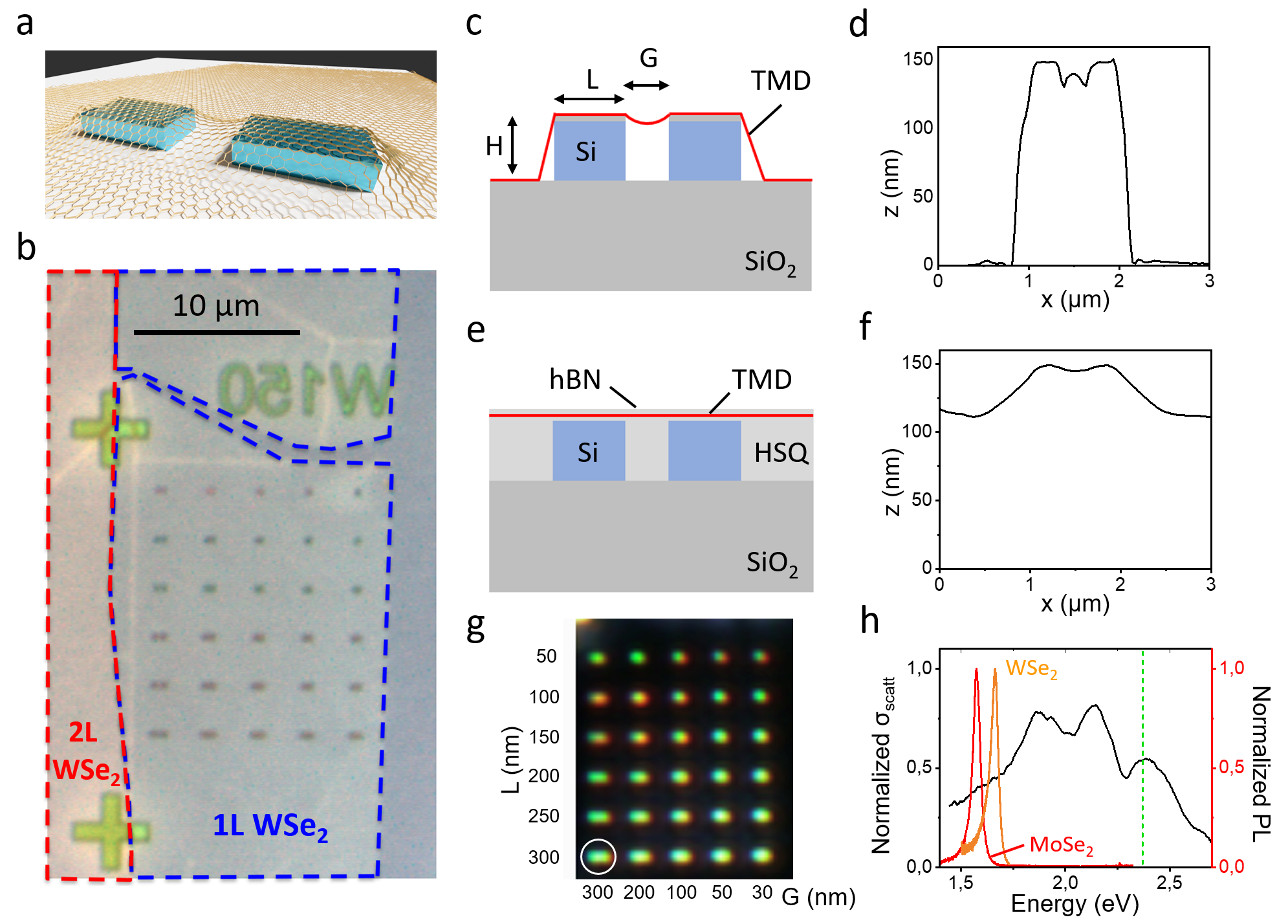}
\caption{\textbf{Nanoresonator arrays for non-planarized and planarized samples.}\textbf{a)} Schematic of a bare monolayer on a non-planarized Si nanoresonator, which is referred to as a dimer and consists of 2 rectangles parallel to each other, separated by a gap $G$. \textbf{b)} Bright field optical image of the sample. Array of rectangular Si dimers with constant width $W=150~nm$ but varying length ($L$) and gap ($G$) covered by a single exfoliated WSe$_2$ ML. \textbf{c)} Illustration of the TMD layer configuration when deposited on non-flattened Si antennas. \textbf{d)}  Measured AFM profiles for configuration shown in c). \textbf{e)} Illustration of the TMD layer configuration when deposited on planarized Si nanoantennas for the same dimers shown in b). \textbf{f)} Measured AFM profiles for configuration shown in e).  \textbf{g)} Dark-field optical images showing the light scattered by the antennas. \textbf{h)} Normalized dark field intensity, measured for $G = 300$~ nm, $L = 300$~nm dimer without the TMD MLs (bottom left on figure g - white circle). Normalized PL spectra measured on ML WSe$_2$ (orange) and MoSe$_2$ (red) (far from antennas), laser excitation energy appear as a green vertical dashed line.}
\label{fig:Sampledescription}
\end{figure*}
\indent In this work, we experimentally separate the three main mechanisms that govern the photoluminescence emission properties of the coupled dielectric NS-TMD monolayer system at room temperature, namely (i) interactions with broad-band Mie resonances, (ii) strain in the TMD layer and (iii) orientation of the exciton dipole with respect to the detection. \textit{First}, we address the impact of the dipole orientation by comparing on identical non-planarised resonator samples  (see dimer resonators sketched in Fig.~\ref{fig:Sampledescription}a) the emission of WSe$_2$ (in-plane and out-of-plane dipoles) \cite{park2018radiative,zhou2017probing,PhysRevLett.119.047401} with MoSe$_2$ (only in-plane emission) \cite{lu2019magnetic,robert2020measurement}. 
For WSe$_2$ we show a brightening of the global emission by one order of magnitude on the NS. This is accompanied by a change in overall spectral shape, due to a variation of the relative contributions to the emission from the in-plane and out-of-plane dipoles, which emit at slightly different energies. In the PL mapping, we find a strong dependence of the spectral shape on detection spot position on the NS (edge/center/suspended region).
\textit{Second}, we address the impact of strain as we compare TMD monolayer emission from samples on Si nanostructures (height about 100~nm) with samples on the same Si nanostructures surrounded by silica (planarized samples), where the TMD is strain free, compare Figs.~\ref{fig:Sampledescription}c and e. We measure in non-planarized samples a clear redshift of the PL energy (correlated to the TMD phonon shift measured in Raman mapping), showing the contribution to enhanced emission of the tensile strain-induced band gap lowering. Strained MoSe$_2$ behaves differently compared to WSe$_2$ : although we see a PL energy redshift due to strain, the spectra consist only of a main, single peak corresponding to the in-plane dipole in MoSe$_2$ (bright exciton). On planarized samples we observe WSe$_2$ and MoSe$_2$ emission with less PL intensity enhancement (50\%) as compared to the strained case, but with a clear impact of NS geometry on the PL emission pattern and dependence on the excitation laser polarization.
Our experiments show an enhanced brightness for all NS-TMD configurations as compared to the TMD alone, so the Si-NS is beneficial in all cases. \\
\begin{figure*}
\includegraphics*[width=0.82\linewidth]{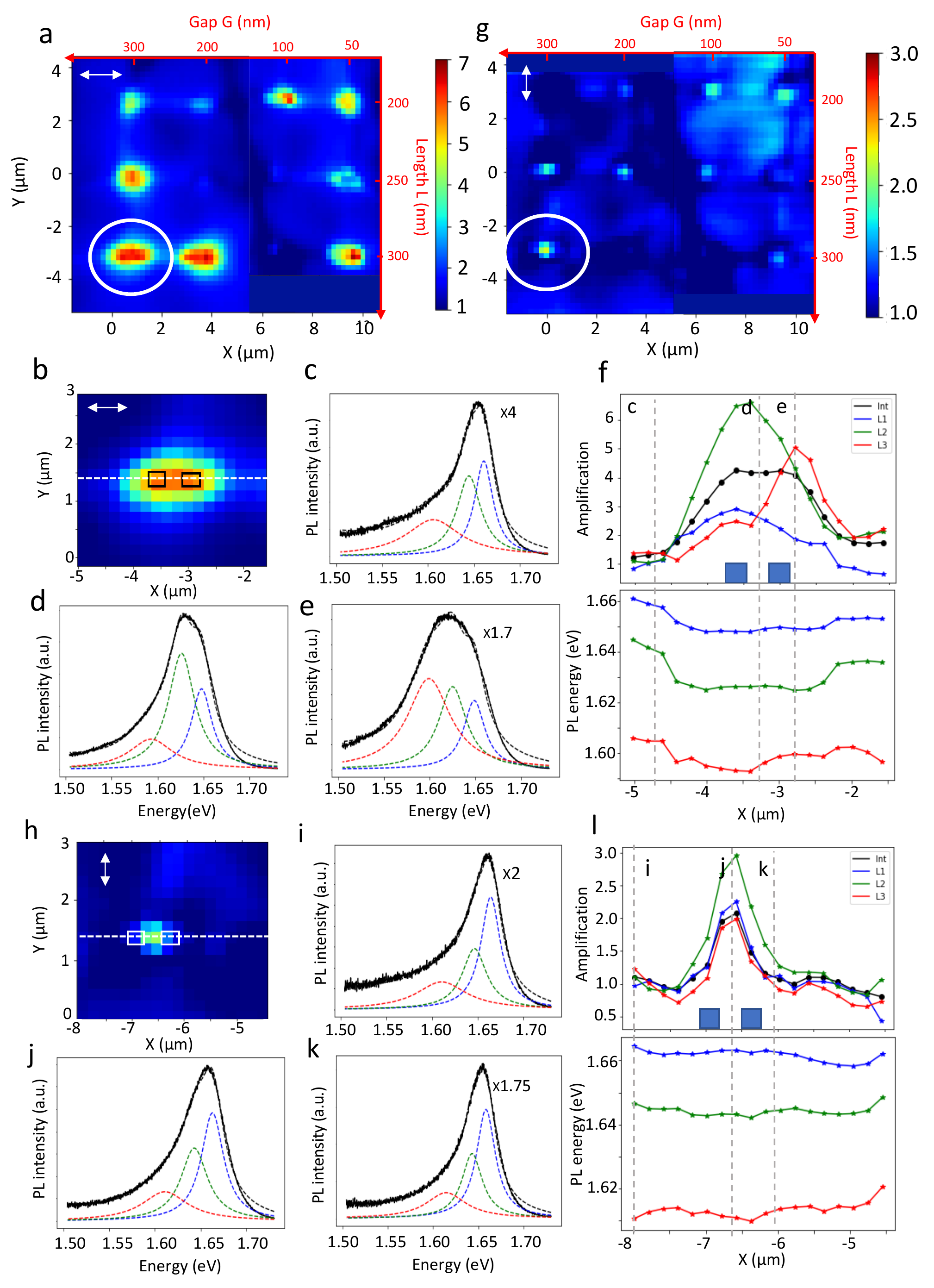}
\caption{\textbf{WSe$_{2}$ monolayer on non-planarized nanoresonators.}\textbf{a)} Normalized WSe$_{2}$ PL color maps of 12 dimer resonators (from $G = 300$~nm, $L = 300$~nm to $G = 50$~nm, $L = 200$~nm), for x-polarized incident light (along the antennas). 
\textbf{b)} Normalized PL color map at the vicinity of resonator $G = 300$~nm, $L = 300$~nm for x-polarized incident light (corresponding to white circle in a)). Dimer position and shape are marked by the black rectangles on the map.  \textbf{c)~d)~e)} show typical PL spectra taken along the white line on figure b)
\textbf{f)} top: Normalized total PL amplitude (black line) and individual Lorentzian fitting component amplitude (red, green, blue) measured across the white line on (b)
; bottom: energy of the individual Lorentzian peaks versus position. The dimer position is shown by the blue square, dashed grey lines mark the positions where the PL spectra shown in panel c) d) e) have been taken.
\textbf{g)} as \textbf{a)} but for y-polarized incident light (perpendicular to the antennas). 
\textbf{h)} Normalized PL color map at the vicinity of resonator $G = 300$~nm, $L = 300$~nm for  y-polarized incident light. Dimer position marked by white rectangles. 
\textbf{i)~j)~k) }show typical PL spectra taken along the white line on figure h). Experimental data appear in black continuous line. Blue, green and red dashed lines are the constituting Lorentzian peak used to fit the PL, the corresponding global fit is shown as black dashed lines. 
\textbf{l)} same as f) but for y-polarized incident light.}
\label{fig:WSe2strained}
\end{figure*}
\textbf{Samples and optical spectroscopy techniques.--- } In this section we describe the two different kinds of nanoresonators used, which show broad resonances covering the wavelength region that corresponds to the MoSe$_2$ and WSe$_2$ monolayer emission and excitation.\\
\indent Si-NS are obtained by a top-down approach in a Silicon on Insulator (SOI) wafer, where a hydrogen silsesquioxane (HSQ) resist is patterned by electron beam lithography followed by reactive ion etching. The thickness of the single crystal overlayer which determines the Si-NS height is \textit{H}=90 nm. After etching and annealing, a SiO$_{2}$ layer (from the HSQ deposition) of about 20 nm is left on top of the Si-NS. We will focus on dimers consisting of two nanorods of rectangular cross section having dimensions of $L \times W \times H$  and separated by a gap \textit{G} (Fig.~\ref{fig:Sampledescription}c). Values of \textit{L} and \textit{G}  are varying across the resonator array from 50 to 300~nm,  for constant \textit{W}=150nm and \textit{H}=90~nm. An array of fabricated structures can be seen in Fig.~\ref{fig:Sampledescription}b. Before TMD monolayer transfer, we perform dark-field scattering experiments to probe the optical Mie resonance bandwidth of the different dimers  \cite{won2019into,kuznetsov2016optically,bidault2019dielectric}. A typical dark-field spectrum is shown in Fig.~\ref{fig:Sampledescription}h, where the measured resonances extend up to the relevant exciton transitions in the TMDs. As it can already be seen in Fig.~\ref{fig:Sampledescription}g by the different colors, the characteristics of the Mie resonances can be varied by changing \textit{L}, \textit{W} and \textit{G} \cite{wiecha2017evolutionary}.\\
\indent Using standard dry-stamping techniques \cite{castellanos2014deterministic} a large monolayer flake of either WSe$_{2}$ or MoSe$_{2}$ is directly transferred on the patterned substrate, covering 30 Si-NS with the same flake in Fig.~\ref{fig:Sampledescription}b. An atomic force microscope (AFM) profile in Fig.~\ref{fig:Sampledescription}d shows that the WSe$_{2}$ ML is acting as a nearly conformal coating, following the Si-NS morphology. These samples are referred to as non-planarized and will be used to investigate the impact of strain and dipole orientation on the PL emission (Fig.~\ref{fig:Sampledescription}c).\\
\indent In order to eliminate strain effects and focus solely on the impact of the Si antenna on the TMD emission, another set of samples was fabricated. 
We followed the same fabrication procedure as for the non-planarized Si-NS and added a planarization step involving HSQ deposition and subsequent annealing and chemical etching \cite{guerfi2015thin}. This leads to Si nanoantennas buried in silica and we refer to this as planarized samples, see Fig.~\ref{fig:Sampledescription}e. The distance between the top Si-NS and the silica smooth surface (see AFM profile Fig.~\ref{fig:Sampledescription}f) is about 30~nm over 1~$\mu$m.  Although the surface is not totally flat, the height variations are smooth enough to avoid strain in the TMD, as confirmed in PL and Raman mappings (see below). The active region is here protected by a 10 nm-thick hexagonal BN layer (high crystal quality, atomically flat \cite{taniguchi2007synthesis}) on top of the TMD ML.\\
\indent Room temperature photoluminescence and Raman experiments are performed using either a Horiba JY Labram HR or a Horiba JY XPlora MV2000 set-up. Hyper-spectral maps are obtained by raster scanning the sample under a laser spot (532~nm excitation wavelength, power of tens of nW) tightly focused through a microscope objective (0.9 numerical aperture - NA). The spectra are obtained in backscattering using the same objective. The high NA allows us to collect mainly emission from in-plane but also to a small extent from out-of plane optical dipoles \cite{brotons2019out,PhysRevLett.119.047401}. Although the diffraction limited spot has a diameter comparable to the dimer, the recorded PL emission map of the sample reveals fine details in lateral resolution due to the high signal to noise ratio. The laser polarization is kept constant and the so-called x (linear polarization along the dimer axis) and y polarization (linear polarization perpendicular the dimer axis) are obtained by rotating the sample by 90$^{\circ}$. Here we define as \textit{the dimer axis} a straight line across the dimer gap that connects the centers of the two Si rectangles.\\
\begin{figure*}
\includegraphics[width=0.95\linewidth]{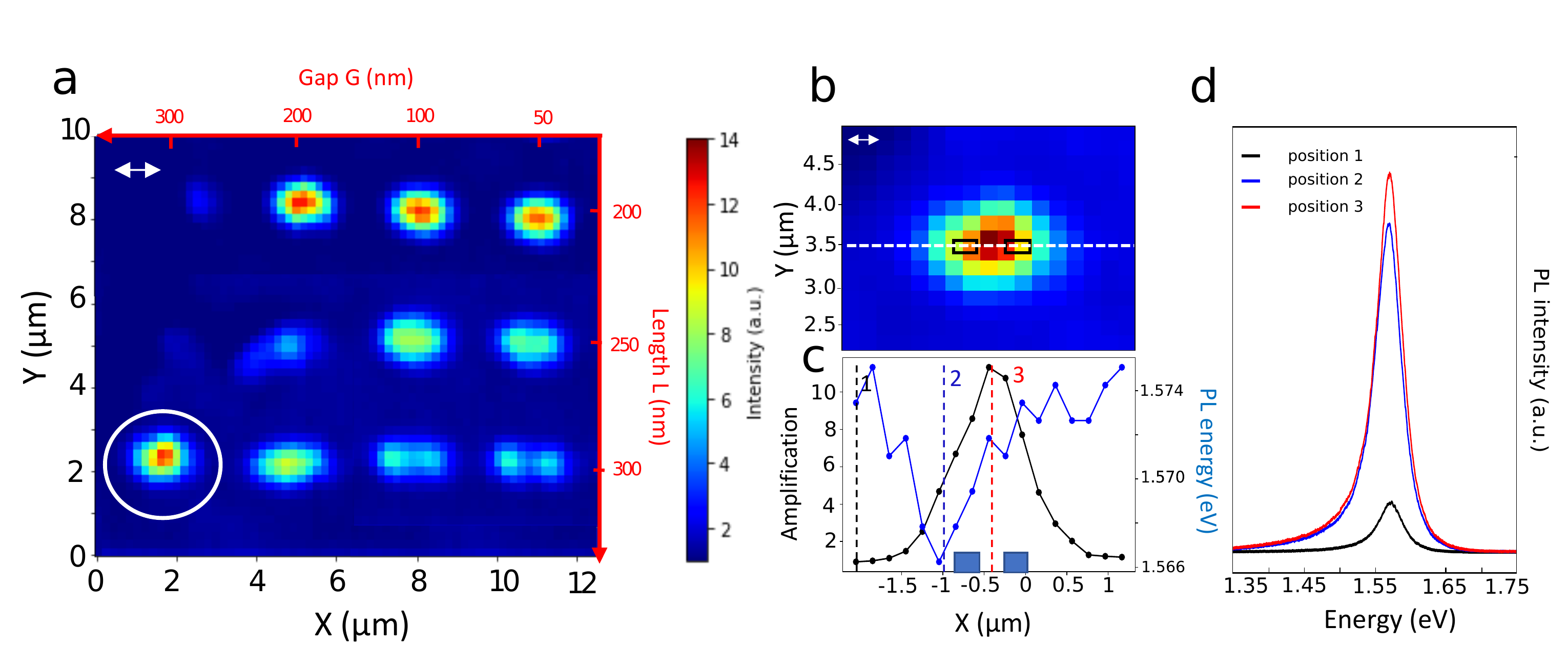}
\caption{\textbf{MoSe$_{2}$ monolayer on non-planarized nanoresonators.} \textbf{a)} Normalized  MoSe$_{2}$ PL color maps of 12 dimer resonators (same nanoresonators as in Fig.~ \ref{fig:WSe2strained}a), for x-polarized on non-planarized nanoresonators. \textbf{b)} Normalized PL color map at the vicinity of nanoresonator $G = 300$~nm, $L = 300$~nm for x-polarized (corresponding to white circle in a)). Dimer positions and shape is marked by the back rectangles on the map. \textbf{c)} Normalized PL integrated amplitude (black line) and energy of the PL maximum is (blue line) versus position, cross section taken along the white dashed line on (b). The dimer position is marked by the blue squares. \textbf{d)} Typical PL spectra taken at 3 different positions marked by matching color dashed vertical lines in (b). }
\label{fig:MoSe2strained}
\end{figure*}

\begin{figure*}
\includegraphics[width=0.95\linewidth]{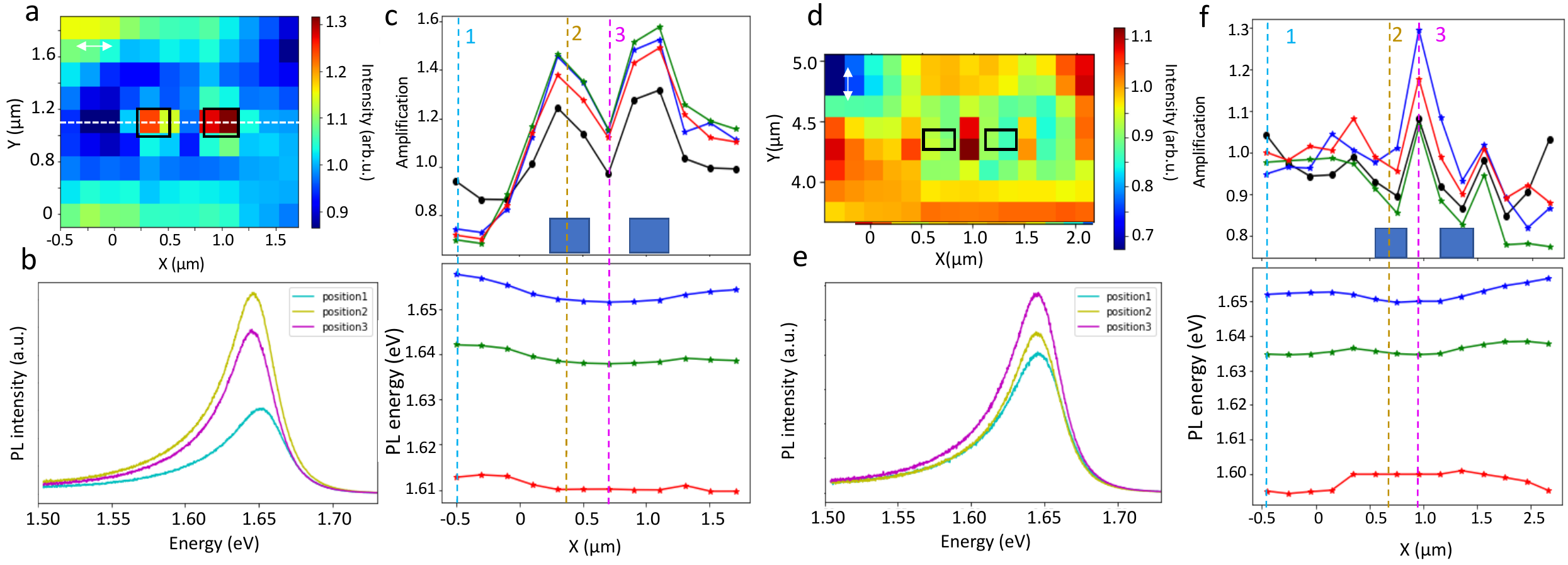}
\caption{\textbf{Strain free WSe$_{2}$ monolayer on planarized nanoresonators.} \textbf{a)} Normalized PL images of  WSe$_{2}$ monolayer deposited on $G= 300$~nm, $L = 300~$nm planarized nanoresonator for x-polarized laser excitation. Dimer position shown as black rectangles. \textbf{b)} Typical PL spectra taken along the white ligne on panel (a). The PL spectra are fitted using three peaks are detailed in Fig.~\ref{fig:WSe2strained}.  \textbf{c)} Top: Normalized total PL amplitude (black line) and individual Lorentzian fitting component amplitude (red, green, blue) measured across the white line on (a). The dimer position is marked by the blue scares. Bottom: Energy of the individual Lorentzian peaks versus position along the white line on panel (a). The positions where the PL spectra have been taken appear as dashed line, with matching color.\textbf{ d), e), f)} same as a), b) and c) but for y-polarized incident light.}
\label{fig:PlanarPL}
\end{figure*}

\begin{figure*}
\includegraphics[width=0.85\linewidth]{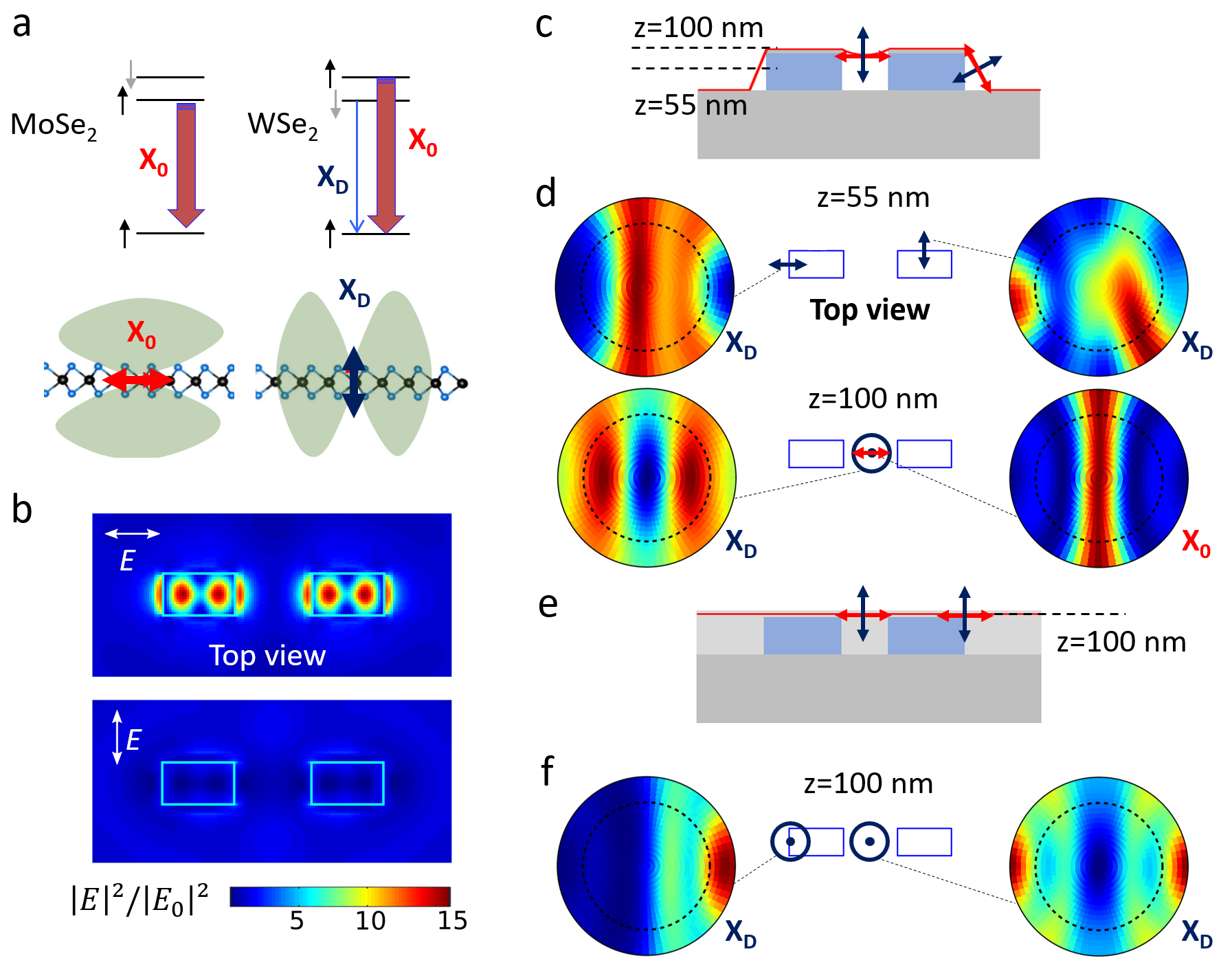}
\caption{\textbf{Numerical simulations of electric field amplitudes and optical dipole emission.} \textbf{a)} Top: schematics of the energy levels in MoSe$_2$ (left) and WSe$_2$ (right). Bottom: orientation of the in-plane neutral exciton X$_o$ (red arrow) and the out-of-plane dark exciton X$_D$ (blue arrow) in the x-z plane with respect to the TMD layer. The gray shaded areas represent the emission lobes of the dipoles. \textbf{b)} Near field intensity in the TMD layer of the non-planarized system for two polarizations parallel and perpendicular to the dimer axis computed in the x-y plane for an incident plane wave at the excitation wavelength $\lambda=532$~nm. \textbf{c)} Geometry of the TMD on a non-planarized dimer in the x-z plane, direction of X$_o$ (red) and X$_D$ (dark blue) are indicated by arrows. \textbf{d)} Emission patterns of a dipole emitting at 750~nm and placed at different positions and heights (z) with respect to the non-planarized dimer shown in (b). The relative positions of the Si-NS and the dipoles are displayed in the x-y plane. The dotted circle corresponds to the 0.9 NA of the microscope objective, we hence detect emission from inside this circle. The dipole can be assimilated to an in-plane exciton (red arrow) or out-of-plane exciton (dark blue arrow) depending on its orientation and position. The surrounding refractive index is n=~1. \textbf{e)} Geometry of the TMD on a planarized dimer in the x-z plane. f) Emission patterns of a dipole emitting at 750~nm and placed at different positions and z=100~nm with respect to the planarized dimer shown in (e). The surrounding refractive index is n=1.45 (HSQ refractive index).
}
\label{fig:Simulation}
\end{figure*}

\indent \textbf{WSe$_{2}$ monolayer on non-planarized nanoresonators.---}
In Fig.~\ref{fig:WSe2strained}a we plot a normalized, spectrally integrated PL map of non-planarized nanoresonators covered by one large WSe$_2$ monolayer. The nanoresonator positions can be seen directly on the map, PL away from the resonators is very weak (uniform dark blue area).
To allow for a direct comparison of the results in Figs.~\ref{fig:WSe2strained}, ~\ref{fig:MoSe2strained},~\ref{fig:PlanarPL} we show spectra from the Si dimer nanoresonator with the same dimensions in all cases of gap $G=300~$nm and length $L=300~nm$ (white circle in Figs.~\ref{fig:WSe2strained}a and g, ~\ref{fig:MoSe2strained}a). This dimer has broad optical Mie resonances at the emission as well as the laser excitation energy (See Fig.~\ref{fig:Sampledescription}h). Strain will clearly play an important role for non-planarized nanoresonators and similar structures are used to generate localized excitons (0D) at low temperature for quantum optics \cite{branny2017deterministic,palacios2017large}. Note that we avoid exciton localization in our room temperature measurements as the thermal energy of excitons is larger than the localization energy. \\
\indent In our experiment in Fig.~\ref{fig:WSe2strained}a the excitation laser is polarized along the main axis of the dimer (x), Fig. \ref{fig:WSe2strained}g shows identical measurements, but carried out with laser polarization perpendicular to the dimer (y) (white arrows indicating the electric field in the figures). 
We first comment on x-polarized excitation. 
The WSe$_2$ PL intensity is substantially enhanced by  one order of magnitude on top of the nanoresonators. 
In addition to intensity changes, we clearly observe that the spectra recorded on top and next to the nanoresonator shown in Fig. \ref{fig:WSe2strained}b are very different, compare panels Fig. \ref{fig:WSe2strained}c,d and e. The PL emission is fitted using three peaks, in agreement with the different transitions identified in the literature for WSe$_{2}$ ML even at room temperature : the free exciton at about 1.66 eV (745 nm) (dashed blue line), the charged excitons (trions) (dashed green line) at about 1.64 eV (756 nm), and the dark (out-of-plane) exciton near 1.61 eV (770 nm) (dashed red line) \cite{park2018radiative,zhou2017probing,PhysRevLett.119.047401}. Please note that we have identified the respective dipole orientation of the dark and bright excitons previously by imaging the Fourier plane of the microscope objective in experiments on ML WSe$_2$ \cite{PhysRevLett.119.047401}.  The relative weight of the different peaks strongly varies as a function of the position along the map, especially around and above the Si-NS (see Fig.\ref{fig:WSe2strained}f top panel). \\
\indent The impact of strain on the optical transitions is revealed in Fig. \ref{fig:WSe2strained}f bottom panel, where we plot the emission energy of the three PL components as a function of position. 
A shift of the PL emission to lower energy for all three transitions is a clear confirmation for tensile strain in the layer \cite{schmidt2016reversible}. The same conclusion can be independently drawn from the Raman maps, presented in the supplement. The enhancement observed for emission on the nanoresonators benefits therefore from the combined effect of strain (local lowering of the band gap that results in funneling of excitons) and optical resonances, which we aim to separate in the next section. Interestingly, as one can see in Fig. \ref{fig:WSe2strained} (f) top panel, the emission linked to the dark exciton (red line), which has an optical dipole out-of the monolayer plane, shows strongest enhancement on the sample edges. Here we suggest that, as the monolayer is folded with respect to the microscope objective axis (by 30$^{\circ}$ to 40$^{\circ}$, as given by AFM profile), more light emitted by the out-of-plane dipole  is detected. These arguments are developed in more detail  in the discussion section. In addition, enhanced emission from excitons with out-of-plane dipoles might be due to lowering of the crystal symmetry \cite{branny2017deterministic,luo2020exciton}. The impact of strain for WSe$_{2}$ MLs on GaP dimers is discussed in references \cite{sortino2020dielectric, sortino2019enhanced}.\\
\indent In Fig.~\ref{fig:WSe2strained}g we plot the same intensity map as in Fig.~\ref{fig:WSe2strained}a, but the excitation laser polarization is now perpendicular to the dimer axis. We observe clear differences: the global enhancement on the NS is now a factor of 3, as compared to roughly 8 for the other laser polarization, and the spatial region around the NS position that shows enhancement is smaller. The maximum enhancement is localized in the gap between the two rectangular rods constituting the dimer, as can be seen in Fig. \ref{fig:WSe2strained}h.  In the dimer gap the monolayer is suspended, and therefore presumably less strained than on the dimer edges. As a result, the PL emission energy does not change as a function of position as we scan across the gap, see Fig.\ref{fig:WSe2strained}l bottom panel.\\ 
\indent Our measurements also show, in contrast to laser excitation \textit{parallel} to the dimer axis,  the global shape of the PL spectra is not affected in the \textit{perpendicular} case (Fig.~\ref{fig:WSe2strained}i,j,k). The relative amplitude of the different PL components remains constant when scanning over an antenna (Fig.~\ref{fig:WSe2strained}l top panel). We suggest in the discussion section below, that this striking dependence on laser excitation polarization (compare Figs.~\ref{fig:WSe2strained}a,b with Figs.~\ref{fig:WSe2strained}g,h) is linked to an optical antenna effect.\\
\indent \textbf{MoSe$_{2}$ monolayer on non-planarized nanoresonators.---}
For WSe$_2$ we observe on several Si-NS dimers a strong PL intensity enhancement above each element of the dimer, showing amplification as high as 8 as compared to the bare monolayer. We now investigate emission from a MoSe$_2$ monolayer deposited on an identical set of resonators, see Fig.~\ref{fig:MoSe2strained}a. We choose this material for comparison as the emission wavelength is close to WSe$_2$ and can be covered by the broadband resonances of the Si-NS as shown in Fig.~\ref{fig:Sampledescription}. An additional advantage for this material is the comparatively simple emission spectrum as the dark, z-polarized states (optical dipole out-of-plane) are in energy above the bright states \cite{lu2019magnetic,robert2020measurement} and therefore have only very small impact on PL emission here, see spectra in Fig.~\ref{fig:MoSe2strained}d and scheme in Fig.~\ref{fig:Simulation}a. \\
In Fig.~\ref{fig:MoSe2strained}a we plot the PL emission map above the resonators and we see strong PL emission enhancemement (up to 14~$\times$) on several resonators covered by the ML flake. PL emission next to the resonators is week (uniform dark area in the map).  The most striking observations are: First, a strong enhancement of the PL emission intensity above the Si-NS, see Fig.~\ref{fig:MoSe2strained}b.  Second, in contrast to our WSe$_2$ results, the PL emission for MoSe$_2$ does not change shape, as can be seen in Fig.~\ref{fig:MoSe2strained}d. The PL band could be fitted using a single peak attributed to the neutral bright exciton. Thus, the red shift of the PL energy shown in Fig.~\ref{fig:MoSe2strained}c is deduced from the maximum of the measured PL peak. As in the case of WSe$_2$, this is a signature of tensile strain but resulting in a smaller energy shift, in qualitative agreement with previous studies on strain tuning \cite{island2016precise}. Although due to the manual stamping method the coupling efficiency of the TMD monolayer might not be uniform across the nanoresonator array, we observe for the investigated MoSe$_2$ monolayer very similar enhancement for several nanoresonators analyzed, as can be seen in the map in Fig.~\ref{fig:MoSe2strained}a. Interestingly for dimers with narrower gaps two emission lobes appear, which might be due to an antenna effect that redirects the light at the dimer edge towards our microscope objective. The simple PL shape makes MoSe$_2$ a suitable material for analyzing more complex effects related to Mie resonances, that might be obscured by the multiple transitions in WSe$_2$. Please note that MoSe$_2$ monolayers have in general comparatively weaker emission than WSe$_2$ monolayers at room temperature \cite{wang2015spin,withers2015wse2,zhang2015experimental} and hence a wider margin for increasing emission efficiency.\\ 
\indent Concerning the dependence on laser excitation polarization for MoSe$_{2}$ we observe a stronger emission for excitation along the dimer axis than perpendicular to it. The difference is about 15~\%, see supplement, much less pronounced than for the WSe$_2$ case, which suggests that the contributions from out-of-plane dipole emission plays an important role for the strong polarization difference observed in WSe$_2$. Note that we analyze for both materials the same nanoresonators with the same optical resonances (as shown in Fig.~\ref{fig:Sampledescription}h), so major differences in our comparative studies between MoSe$_2$ and WSe$_2$ emission have their origin in the different bandstructure, sketched in Fig.~\ref{fig:Simulation}a. \\

\indent \textbf{Strain free WSe$_{2}$ monolayer on planarized nanoresonators.---}
In the case of the planarized samples, neither strain-induced Raman shifts of the TMD phonon can be detected (see supplement) nor a systematic redshift of the PL energy, as shown in our measurements in Fig.~\ref{fig:PlanarPL}c and f. The AFM profile in Fig.~\ref{fig:Sampledescription}f shows that the maximum roughness above a Si-NS is limited to about  30~nm extended over 1~$\mu$m. For this strain-free WSe$_2$ monolayer on planarized resonators we make the following observations: We record an increase of the PL of 50\%  due to the presence of Si-NS, this is less than the factor of about 8 achieved on the same nanoresonator but non-planarized, compare Fig.~\ref{fig:WSe2strained}a and Fig.~\ref{fig:PlanarPL}a.  Interestingly, the overall emission spectrum for the planarized sample of WSe$_{2}$ does not change shape but only the amplitude is modified by the antenna, see Fig.~\ref{fig:PlanarPL}b and e. The bright exciton contribution in the PL band is the strongest peak according to our fits and the weight of the different contributions (dark and charged excitons)  are constant when we scan across the Si-NS.  The PL maps in Fig.~\ref{fig:PlanarPL} are dependent on the laser polarization for WSe$_2$, compare panels Fig.~\ref{fig:PlanarPL}a,c with  Fig.~\ref{fig:PlanarPL}d,f. For y-polarized excitation we see a maximum in PL signal over the dimer gap, see Fig.~\ref{fig:PlanarPL}d,f.
Our measurements show that the emission of WSe$_2$ is governed by an optical antenna effect in the planarized samples, 
whereas strain effects play a key role in the nonplanar samples, as pointed out in earlier work on strain WSe$_{2}$ on GaP dimers \cite{sortino2019enhanced}. 
The measured PL enhancement is  up to 50\% of the integrated intensity, which can be further improved. Our measurements on planarized MoSe$_{2}$ samples gave very similar results. \\
\indent In future experiments the Si-NS shape and dimensions can be optimized in order to reach higher electric field amplitudes. It is important to take into account that, after HSQ planarization and hBN deposition, the dielectric contrast between Si-NS and the surrounding is lower than for non-planarized resonators. This lowers light confinement and needs to be compensated by optimizing the antenna design.\\

\indent \textbf{Numerical simulations and discussion.---}
Our measurements of the PL emission of ML WSe$_2$ and MoSe$_2$ coupled to non-planarized and planarized nanoresonators revealed for all cases an enhancement of the PL emission as compared to the bare monolayer. In the discussion below we suggest different mechanisms at the origin of this enhancement taking into account the polarization-dependent near field at the excitation energy and the interaction of dipole in-plane and out-of the ML plane with the Mie resonances of the dimer structures at the emission energy.\\
\indent The Si-NS is optically resonant at the excitation wavelength. We first simulate the spatial distribution of the electric near-field intensity in the plane of the TMD monolayer (see Fig.~\ref{fig:Simulation}c) with the finite-difference time-domain (FDTD) method  \cite{taflove_computational_2010}, 
using an open-source software package \cite{oskooi_meep_2010}, see results in Fig.~\ref{fig:Simulation}b. For laser excitation parallel to the dimer axis our simulation shows a stronger near-field enhancement than in the case where the polarization is perpendicular to the dimer axis. The near-field enhancement is localized above and around the Si blocks in the parallel polarization case. This can be seen in particular along the edges (y-oriented) perpendicular to the polarization. In contrast, there is almost no enhancement for the perpendicular case (even lower field above each  block of the Si dimer).
This behavior, relevant at the laser excitation wavelength, is in qualitative agreement with the PL maps of the chosen example described previously, compare Fig.~\ref{fig:WSe2strained}a and Fig.~\ref{fig:WSe2strained}g. This indicates that optical Mie resonances contribute to PL enhancement. Nevertheless Mie resonances alone cannot explain the differences in the enhancement and the polarization dependence observed in the four configurations (ML WSe$_2$ and ML MoSe$_2$ on non-planarized and planarized Si-NS, respectively), otherwise the observations for MoSe$_2$ and WSe$_2$ for the polarization dependence would have been identical.\\ 
\indent To go further, we study the antenna effect induced by Si-NS on the emission of the two kinds of exciton dipoles (the bright  exciton X$_o$ and the dark  exciton X$_D$) \cite{park2018radiative,zhou2017probing,PhysRevLett.119.047401}. In a first approximation we consider a single oscillating dipole with orthogonal orientation depending on the nature of the exciton (in-plane X$_o$ and out-of-plane X$_D$ - see Fig.~\ref{fig:Simulation}a). 
Considering the non-planarized and planarized geometries (Fig.~\ref{fig:Simulation}c and e), we compute the intensity of the electric field collected in backscattering geometry through the microscope objective NA for several key locations of the emitting dipoles. The electric field radiated by the dipolar source in presence of a Si-NS is obtained by the Green Dyadic Method \cite{girard_near_2005} using the pyGDM toolkit for electrodynamics simulations \cite{wiecha_pygdmpython_2018, wiecha_enhancement_2019}. In Fig.~\ref{fig:Simulation}d-f are shown the emission patterns of different dipoles emitting at 750~nm (approximately the average of the bright and dark exciton wavelengths).\\
\indent In the case of the upper left part of Fig.~\ref{fig:Simulation}d, when the ML is nearly conformal to the Si-NS i.e. following its shape, the intensity measured through the microscope objective (represented by the dotted circle) of an out-of-plane dipole (X$_D$) positioned perpendicular and at mid-height of an y-oriented edge is very high. In addition, the collected intensity originating from the in-plane dipoles is negligible. \\
\indent In contrast, if the dipole is positioned at mid-height of an exterior x-oriented edge, corresponding to the upper right part of Fig.~\ref{fig:Simulation}d, the collected intensity drops. We believe that this antenna effect plays an important role in the increase of the measured signal. It adds to the polarization-dependent enhancement presented in Fig.~\ref{fig:Simulation}b, which leads to the creation of more excitons (including X$_D$ ones) in the regions of high electric field. Thus, the combination of these two fundamental mechanisms leads to the boost in the PL enhancement observed for WSe$_{2}$ on nonplanar Si-NS, only for a parallel polarization of the laser with respect to the Si dimer axis. \\ \indent 
In the gap of the Si dimer (lower part of Fig.~\ref{fig:Simulation}d), the integrated intensity corresponding to X$_D$ emission is approximately the same as the one due to X$_o$. Both are contributing, thus leading again to the signal increase. Compared to the emission pattern of the dipole in a TMD monolayer alone hosting X$_D$ (WSe$_{2}$) (See Fig.~\ref{fig:Simulation}a), it is clear that the relative contributions of X$_D$ and X$_o$ are very different as a function of the emitting dipole position in the case of the non-planar sample. Our simple model is already sufficient to understand the underlying physics leading to the strong PL enhancement when the TMD hosts out-of-plane excitons. For instance, since this cumulative effect mainly occurs along the y-oriented edges for a parallel polarization, it helps explaining the extension of the PL patterns along the x-axis well beyond the dimer limits in the maps of Fig.~\ref{fig:WSe2strained}. For the same polarization configuration, it is also in good agreement with the PL band shape modification related to the strong enhancement of the fitted (red) peak associated to X$_D$ at the y-oriented edges of the Si-NS in the spectra of Fig.~\ref{fig:WSe2strained}.\\
\indent On the other hand, for the planarized sample, the dark exciton contribution is always very minor and the PL band is dominated by the bright exciton contribution, compare Fig.~\ref{fig:PlanarPL}e and Fig.~ \ref{fig:WSe2strained}e. The radiation pattern of the X$_D$ is weakly redirected by the antenna in the angular window defined by the microscope objective NA, neither when it is positioned along the edges, nor in the gap region (Fig.~\ref{fig:Simulation}f). As a consequence, no enhancement associated to the X$_D$ is expected. It is in good agreement with the experimental observation where this dark exciton is intrinsically eliminated (MoSe$_{2}$ samples) or does experience no enhancement (WSe$_{2}$ on planarized  samples). In both cases, the polarization-dependent enhancement, resulting from a pure optical effect associated to the Mie resonances in these Si-NS, is about 50\%, far from the roughly 10$\times$ enhancement in PL from WSe$_{2}$ on nonplanar resonators.\\ 
\indent We unveil here three intertwined mechanisms that contribute to the PL enhancement. The Mie-assisted enhancement of the dark exciton contribution plays therefore an important role in the global PL response of the WSe$_{2}$- resonator coupled systems. Interestingly, its contribution seems to be of the same order of magnitude as the strain one. The amplitude of the latter can be estimated with MoSe$_{2}$ on non-planar samples. Indeed, about 15 \% variation of the PL is observed as the polarization is rotated, indicating a very strong contribution to the global factor  PL enhancement (14~$\times$) originates from the strain generated by the Si nanoblocks in this sample.\\ 
\indent Finally, as expected for non-optimized resonators, the pure optical effect associated to Mie resonances observed in planarized samples in Fig.~\ref{fig:PlanarPL} (no strain, no dark exciton) is relatively weak compared to the two other mechanisms. It should easily be further improved by optimizing the geometry of the nanoresonators.\\
\indent \textbf{In conclusion}, we have shown strong PL enhancement for WSe$_{2}$ and MoSe$_{2}$ on non-planarized nanoresonators due to the combined effect of amplifying contributions from out-of plane dipole emission, strain and pure electric field amplification. On planarized nanoresonators we show that the TMD monolayer is strain free and the PL emission is only enhanced by the Mie resonances of the Si dimer. In addition to presence and absence of strain in the active medium for non-planarized and planarized samples, respectively, also the dielectric contrast plays a role. For non-planarized resonators dielectric contrast is high, whereas it is lower for planarized samples, which demands further work on planarization techniques such as the use of low refractive index dielectric around the silicon nanostructures, and a very small thickness to position the TMD as close as possible to the Si nanoantenna. Finally the optimization of the nanoantenna design will be essential to control strain, resonances at both the excitation and emission wavelength, and directivity of the emitted light. WSe$_2$ is a very promising material as the in-plane and out-of-plane exciton emission can be controlled independently to increase the global emission yield. MoSe$_{2}$ shows also strong PL emission enhancement and a simple emission spectrum for targeted interaction with Mie resonances. \\
\indent \textbf{Acknowledgements.---} LPCNO acknowledges funding from ANR 2D-vdW-Spin, ANR MagicValley, ITN 4PHOTON No. 721394 and the Institut Universitaire de France. CEMES, LAAS, LETI and LPCNO acknowledge funding from ANR HiLight (ANR-19-CE24-0020-01). This study has been partially supported through the EUR grant NanoX  2D Light (ANR-17-EURE-0009) in the framework of the Programme des Investissements d'Avenir. 
K.W. and T.T. acknowledge support from the Elemental Strategy Initiative
conducted by the MEXT, Japan ,Grant Number JPMXP0112101001,  JSPS
KAKENHI Grant Number JP20H00354 and the CREST(JPMJCR15F3), JST.
 CEMES team acknowledges the Calmip computing facility (grant P12167).


\end{document}